\documentclass[english,aps,prl,reprint,superscriptaddress,numerical,showpacs,floatfix,nofootinbib,twocolumn]{revtex4-1}
\usepackage[T1]{fontenc}
\usepackage[latin9]{inputenc}
\usepackage[pdftex]{color}
\usepackage{babel}
\usepackage{amssymb}
\usepackage[pdftex]{graphicx}
\usepackage{tabularx}
\usepackage{footnote}
\usepackage[unicode=true,pdfusetitle,
 bookmarks=true,bookmarksnumbered=false,bookmarksopen=false,
 breaklinks=false,pdfborder={0 0 1},backref=false,colorlinks=true]
 {hyperref}
\hypersetup{urlcolor=blue, citecolor=blue}

\usepackage{dcolumn}
\newcolumntype{d}{D{.}{.}{1} }



\begin{document}

\title{Breakdown of the Isobaric Multiplet Mass Equation for the A = 20 and 21 Multiplets}

\author{A. T. Gallant}
\email{agallant@triumf.ca}
\affiliation{TRIUMF, 4004 Wesbrook Mall, Vancouver, British Columbia, V6T 2A3 Canada}
\affiliation{Department of Physics and Astronomy, University of British Columbia, Vancouver, British Columbia, V6T 1Z1 Canada}
\author{M. Brodeur}
\affiliation{Department of Physics, University of Notre Dame, Notre Dame, Indiana 46556 USA}
\author{C. Andreoiu}
\affiliation{Departmet of Chemistry, Simon Fraser University, Burnaby, British Columbia, V5A 1S6 Canada}
\author{A. Bader}
\affiliation{TRIUMF, 4004 Wesbrook Mall, Vancouver, British Columbia, V6T 2A3 Canada}
\affiliation{\'{E}cole des Mines de Nantes, La Chantrerie, 4, rue Alfred Kastler, B.P. 20722 - F-44307 NANTES Cedex 3}
\author{A. Chaudhuri}
\altaffiliation[Current Address: ]{Rare Isotope Science Project (RISP), 309 Institute for Basic Science, Daejeon 305-811, Republic}
\affiliation{TRIUMF, 4004 Wesbrook Mall, Vancouver, British Columbia, V6T 2A3 Canada}
\author{U. Chowdhury}
\affiliation{TRIUMF, 4004 Wesbrook Mall, Vancouver, British Columbia, V6T 2A3 Canada}
\affiliation{Department of Physics and Astronomy, University of Manitoba, Winnipeg, Manitoba, R3T 2N2 Canada}
\author{A. Grossheim}
\affiliation{TRIUMF, 4004 Wesbrook Mall, Vancouver, British Columbia, V6T 2A3 Canada}
\author{R. Klawitter}
\affiliation{TRIUMF, 4004 Wesbrook Mall, Vancouver, British Columbia, V6T 2A3 Canada}
\affiliation{Max-Planck-Institut f\"{u}r Kernphysik, Saupfercheckweg 1, D-69117 Heidelberg, Germany}
\author{A. A. Kwiatkowski}
\affiliation{TRIUMF, 4004 Wesbrook Mall, Vancouver, British Columbia, V6T 2A3 Canada}
\author{K. G. Leach}
\affiliation{TRIUMF, 4004 Wesbrook Mall, Vancouver, British Columbia, V6T 2A3 Canada}
\affiliation{Departmet of Chemistry, Simon Fraser University, Burnaby, British Columbia, V5A 1S6 Canada}
\author{A. Lennarz}
\affiliation{TRIUMF, 4004 Wesbrook Mall, Vancouver, British Columbia, V6T 2A3 Canada}
\affiliation{Institut f\"{u}r Kernphysik, Westf\"{a}lische Wilhelms-Universit\"{a}t, D-48149 M\"{u}nster, Germany}
\author{T. D. Macdonald}
\affiliation{TRIUMF, 4004 Wesbrook Mall, Vancouver, British Columbia, V6T 2A3 Canada}
\affiliation{Department of Physics and Astronomy, University of British Columbia, Vancouver, British Columbia, V6T 1Z1 Canada}
\author{B. E. Schultz}
\affiliation{TRIUMF, 4004 Wesbrook Mall, Vancouver, British Columbia, V6T 2A3 Canada}
\author{J. Lassen}
\affiliation{TRIUMF, 4004 Wesbrook Mall, Vancouver, British Columbia, V6T 2A3 Canada}
\affiliation{Department of Physics and Astronomy, University of Manitoba, Winnipeg, Manitoba, R3T 2N2 Canada}
\author{H. Heggen}
\affiliation{TRIUMF, 4004 Wesbrook Mall, Vancouver, British Columbia, V6T 2A3 Canada}
\author{S. Raeder}
\affiliation{TRIUMF, 4004 Wesbrook Mall, Vancouver, British Columbia, V6T 2A3 Canada}
\author{A. Teigelh\"{o}fer}
\affiliation{TRIUMF, 4004 Wesbrook Mall, Vancouver, British Columbia, V6T 2A3 Canada}
\affiliation{Department of Physics and Astronomy, University of Manitoba, Winnipeg, Manitoba, R3T 2N2 Canada}
\author{B. A. Brown}
\affiliation{Department of Physics and Astronomy and National Superconducting Cyclotron Laboratory, Michigan State University, East Lansing, Michigan 48824-1321, USA}
\author{A. Magilligan}
\affiliation{Department of Physics, Florida State University, Tallahassee, Florida 32306, USA}
\author{J. D. Holt}
\altaffiliation[Current Address: ]{TRIUMF, 4004 Wesbrook Mall, Vancouver, British Columbia, V6T 2A3 Canada}
\affiliation{Institut f\"ur Kernphysik, Technische Universit\"at
Darmstadt, 64289 Darmstadt, Germany}
\affiliation{ExtreMe Matter Institute EMMI, GSI Helmholtzzentrum f\"ur
Schwerionenforschung GmbH, 64291 Darmstadt, Germany}
\affiliation{Department of Physics and Astronomy and National Superconducting Cyclotron Laboratory, Michigan State University, East Lansing, Michigan 48824-1321, USA}
\author{J.\ Men\'{e}ndez}
\affiliation{Institut f\"ur Kernphysik, Technische Universit\"at
Darmstadt, 64289 Darmstadt, Germany}
\affiliation{ExtreMe Matter Institute EMMI, GSI Helmholtzzentrum f\"ur
Schwerionenforschung GmbH, 64291 Darmstadt, Germany}
\author{J.\ Simonis}
\affiliation{Institut f\"ur Kernphysik, Technische Universit\"at
Darmstadt, 64289 Darmstadt, Germany}
\affiliation{ExtreMe Matter Institute EMMI, GSI Helmholtzzentrum f\"ur
Schwerionenforschung GmbH, 64291 Darmstadt, Germany}
\author{A. Schwenk}
\affiliation{ExtreMe Matter Institute EMMI, GSI Helmholtzzentrum f\"ur
Schwerionenforschung GmbH, 64291 Darmstadt, Germany}
\affiliation{Institut f\"ur Kernphysik, Technische Universit\"at
Darmstadt, 64289 Darmstadt, Germany}
\author{J. Dilling}
\affiliation{TRIUMF, 4004 Wesbrook Mall, Vancouver, British Columbia, V6T 2A3 Canada}
\affiliation{Department of Physics and Astronomy, University of British Columbia, Vancouver, British Columbia, V6T 1Z1 Canada}

\begin{abstract}
\noindent
Using the Penning trap mass spectrometer TITAN, we performed the first direct mass measurements of $^{20,21}$Mg, isotopes that are the most proton-rich members of the $A=20$ and $A=21$ isospin multiplets. 
These measurements were possible through the use of a unique ion-guide laser ion source, a development that suppressed isobaric contamination by six orders of magnitude.
Compared to the latest atomic mass evaluation, we find that the mass of $^{21}$Mg is in good agreement but that the mass of $^{20}$Mg deviates by 3$\sigma$.
These measurements reduce the uncertainties in the masses of $^{20,21}$Mg by 15 and 22 times, respectively, resulting in a significant departure from the expected behavior of the isobaric multiplet mass equation in both the $A=20$ and $A=21$ multiplets.
This presents a challenge to shell model calculations using either the isospin non-conserving USDA/B Hamiltonians or isospin non-conserving interactions based on chiral two- and three-nucleon forces.
\end{abstract} 
\date{\today}
\maketitle

The wealth of data obtained from experiments on increasingly exotic nuclei is only possible because of improved radioactive beam production and delivery \cite{2013Blumenfeld}.
Experiments with these improved beams continually challenge existing theories, a challenge often led by precision mass measurements \cite{2006Blaum}.
High-precision atomic mass measurements are crucial in refining nucleosynthesis abundance calculations \cite{2013Wolf,2013VanSchelt}, in deepening our understanding of fundamental aspects of the strong force \cite{2012Gallant,2013Wienholtz,2013Steppenbeck,2013Hammer}, pointing to nuclear structure change \cite{2013Blaum}, and providing signatures of exotic phenomena such as nuclear halo formation \cite{2013Blaum}.
A tool from theory currently confronted by increasingly precise mass values is the isobaric multiplet mass equation (IMME) \cite{2012Zhang}, 
which relates the mass excesses ME of isospin multiplet members as
\begin{equation}
\mathrm{ME}(A,T,T_{z}) = a(A,T) + b(A,T) T_{z} + c(A,T) T_{z}^{2},
\end{equation}
where $A$ and $T$ are the mass number and total isospin of the multiplet, and $a$, $b$, $c$ are coefficients that depend on all quantum numbers except $T_z$.
In the isospin description of nucleons, the proton and neutron belong to a $T = 1/2$ doublet with projections $T_z(n) = 1/2$ and $T_z(p) = -1/2$.
However, isospin is only an approximate symmetry, broken by electromagnetic interactions and the up and down quark mass difference in QCD.
One can see large deviations from the IMME in the $A=9$ $J^{\pi} = 3/2^-$ and $A = 33$ and $35$ $J^{\pi} = 3/2^+$ quartets 
\cite{2012bBrodeur,2007Yazidjian,2013Lam} and in the $A=8$ and $32$ 
\cite{2011Charity,2009Kwiatkowski,2010Kankainen} quintets.
Suggested explanations
for these deviations include isospin mixing and second-order
Coulomb effects, requiring cubic, $dT_{z}^{3}$, or quartic, $eT_{z}^{4}$, terms to be considered.

Many tests of the IMME on proton-rich systems are hindered by a long standing problem: excessive in-beam contamination.
Substantial isobaric background from alkalis and lanthanides often prevents ground-state measurements of exotic nuclei, especially for nuclei produced at low rates.
This applies even for beams produced using element-selective laser ionization in a classical hot-cavity resonant ionization laser ion source.
We have developed a novel technology for on-line laser ion sources that suppresses any background contamination, enabling a substantial number of experiments to proceed.
This new ion-guide laser ion source (IG-LIS) has allowed the first direct mass measurements of the most proton-rich members of the $A = 20$ and $21$ isospin multiplets.
Being the lightest isospin multiplets where all members are stable against particle emission, and the lightest isospin multiplets which can be described within the $d_{5/2}$, $s_{1/2}$, and $d_{3/2}$ orbitals ($sd$ shell), the $A = 20$ and $21$ multiplets provide an excellent test of the IMME.
This can be done experimentally by high-precision mass measurements of $^{20,21}$Mg and theoretically by shell model calculations using either the USDA/B isospin non-conserving (INC) Hamiltonian or interactions based on chiral effective field theory.

\begin{figure}
	\includegraphics[width=8.6cm]{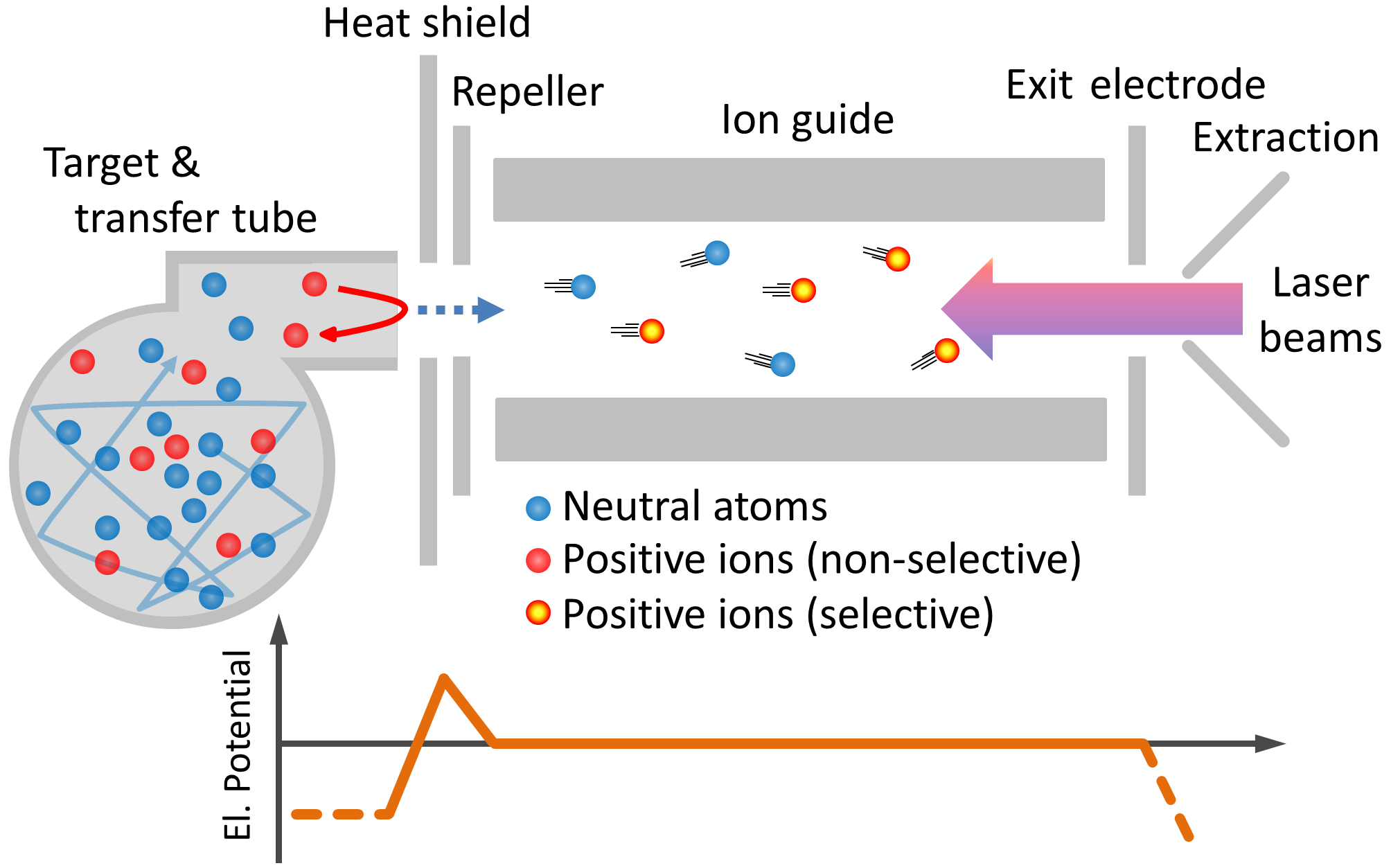}
	\caption{(Color online) A schematic drawing of the ISAC target with the ion-guide laser ion source (IG-LIS). See the text for a detailed description.}
	\label{fig_IG-LIS}
\end{figure}

In a Penning trap, contaminants can be effectively removed if their ratio to the ions of interest remains $\lessapprox100:1$.
It is possible to clean beams with higher levels of contamination using either gas-filled Penning traps \cite{1991Savard} or multi-reflection time-of-flight devices \cite{1990Wollnik,2012Wolf}, but these methods suffer from increased preparation times and  transport losses. 
An alternative method is to suppress contamination from surface-ionized species at the source through the use of an IG-LIS.
The IG-LIS is conceptually similar to the originally proposed ion-source trap \cite{2003Blaum,2013Fink}; however it is much simpler because no trap is formed and no buffer gas is used.
% A section view of the IG-LIS is shown in Fig. \ref{fig_IG-LIS}. 
Figure \ref{fig_IG-LIS} shows a diagram of the IG-LIS. 
The isotope production target is typically operated at temperatures above 1900 K.
The target production volume is coupled via a heated transfer tube to a positively biased radio-frequency quadrupole (RFQ) ion-guide.
A repeller electrode \cite{2013Fink,2013Raeder} reflects surface-ionized species, preventing them from entering the ion-guide volume.
Neutral particles drift into the ion-guide volume, where element-selective resonant laser excitation and ionization creates an isobar free ion beam. 
A square-wave RF field radially confines the laser ionized beam. 
The ion beam is then extracted from the IG-LIS for subsequent mass separation and delivery to the experiment. 
A complete description of the IG-LIS can be found in Ref.~\cite{2013Raeder}.

The IG-LIS concept has been implemented and used online for the first time at TRIUMF's isotope separator and accelerator (ISAC) facility. Beams of $^{20,21}$Mg were produced by bombarding a SiC target with 40 $\mu$A of 480 MeV protons.
Compared to previous SiC targets, we found that the IG-LIS decreased the magnesium yield by 50 times and the sodium background by $10^6$ times.
This is a signal-to-background improvement of better than $10^4$.

\begin{figure}
	\includegraphics[width=8.6cm]{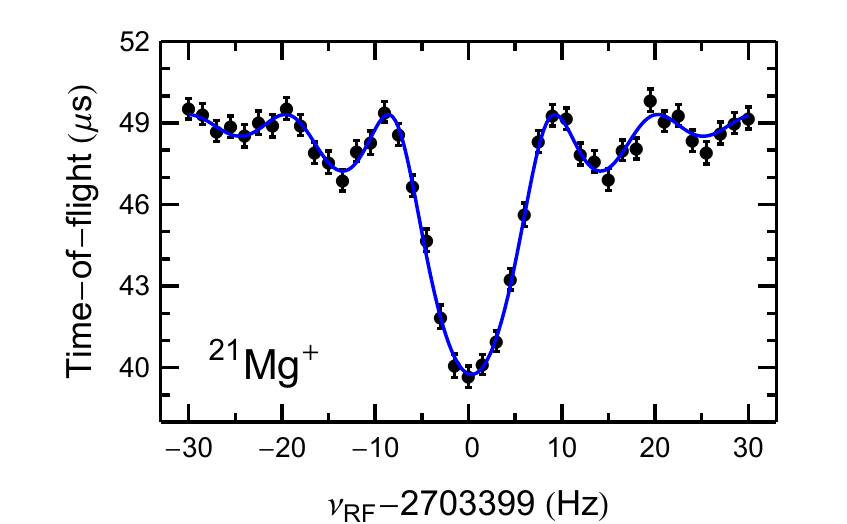}
	\caption{(Color online) Time-of-flight ion cyclotron resonance of $^{21}$Mg with 97 ms excitation time.
	The line is a fit of the theoretical line shape \cite{1995Konig}.}
	\label{fig_20Mgres}
\end{figure}

\begin{table*}[t]
	\caption{Half-lives, number of measurements $N$, detected number of ions $N_{\textrm{ions}}$, excitation times $T_{RF}$, measured frequency ratios and derived mass excesses of $^{20,21}$Mg.
For the frequency ratios, the first two numbers in parentheses are
the statistical and systematic uncertainties, while the number in
brackets is the total uncertainty; otherwise the quoted uncertainty
is the total uncertainty. Half-lives are from the NNDC \cite{2013nndc}.}
\begin{tabular}{c c c c c *{1}{D{.}{.}{3.19}} *{1}{D{.}{.}{6.6}} *{1}{D{.}{.}{12.4}} }
	\hline
	\hline
	Nuclide & $T_{1/2}$ (ms) & $N$ & $N_{\textrm{ions}}$ & $T_{RF} (ms)$ & \multicolumn{1}{c}{Frequency Ratio $r$} &
	\multicolumn{1}{c}{ME (keV)} & \multicolumn{1}{c}{ME$_{\rm{AME2012}}$(keV)} \\
	\hline 
	$^{20}$Mg & 90 (6) & 6 & 1401 & 97 & 0.870765248\ (77)(40)[87] & 17477.7\ (18) & 17559\ (27) \\
	$^{21}$Mg & 122 (2) & 5 & 31369 & 97 & 0.913956913\ (33)(9)[35] & 10903.85\ (74) & 10914\ (16) \\
	\hline 
	\hline
\end{tabular}
\label{table_masses}
\end{table*}

\begin{table}[!ht]
\caption{Extracted IMME parameters for the $A=20$ and $21$ multiplets. Mass excesses are
	taken from \cite{2012Wang} and excitation energies $E_{x}$ from \cite{2013nndc} and
\cite{2004Firestone}, except where noted. Also shown are the $d$ and $e$ coefficients for cubic and quartic fits and the $\chi^2$ values of the fit. Shell model calculation results using the USDA/B plus INC interactions are presented.}
\begin{minipage}{8.6cm}
	\begin{tabularx}{8.6cm}{X X r@{}l r@{}l}
	\hline
	\hline
	Nuclide & $T_{z}$ & \multicolumn{2}{c}{ME(g.s.) (keV)} & \multicolumn{2}{c}{$E_{x}$ (keV)} \\
	\hline
	\multicolumn{6}{l}{$A$ = 20, $J^{\pi} = 0^{+}$, $T = 2$} \\
	$^{20}$O  & +2  & 3796&.17 (89)	& 0&.0		\\
	$^{20}$F  & +1  & -17&.45 (3)	& 6519&.0 (30)	\\
	$^{20}$Ne  & 0	  & -7041&.9306 (16)	& 16732&.9 (27)	\\
	$^{20}$Na  & -1  & 6850&.6 (11)	& 6524&.0 (97) \footnotemark[1] \\
	$^{20}$Mg& -2  & 17477&.7 (18) \footnotemark[2] & 0&.0		\\
	\multicolumn{6}{c}{
		\begin{tabular} {c c c c c}
			Ref. & $a$ (keV) & $b$ (keV) & $c$ (keV) & $\chi^2$ \\
			\hline
			This Work & 9689.79 (22) & -3420.57 (50) & 236.83 (61) & 10.2\\
			Ref. \cite{2013Lam} & 9693 (2) & -3438 (4) & 245 (2) & 1.1\\
			\hline
		\end{tabular}
	}\\
	\multicolumn{6}{c}{
		\begin{tabular}{l c c c}
		Fit & $d$ (keV) & $e$ (keV)& $\chi^{2}$\\
		\hline
		% Quadratic & - & - & 10.4\\
		Cubic & $2.8\ (11)$ & - & 3.7\\
		Quartic Only & - & $0.89\ (12)$ & 9.9\\
		Quartic & $5.4\ (17)$ & $-3.5\ (18)$ & -\\
		USDA & $-0.1$ & - &\\
		USDA & - & $-1.7$ &\\
		USDB & $-0.1$ & - &\\
		\end{tabular}
        }\\
	\hline
	\hline
	\multicolumn{6}{l}{$A$ = 21, $J^{\pi} = 5/2^{+}$, $T = 3/2$} \\
	$^{21}$F  & +3/2 & -47&.6 (18)	  & 0&.0		  \\
	$^{21}$Ne  & +1/2 & -5731&.78 (4)  & 8859&.2 (14)  \\ 
	$^{21}$Na  & -1/2 & -2184&.6 (3)  & 8976&.0 (20)	  \\
	$^{21}$Mg  & -3/2 & 10903&.85 (74) \footnotemark[2] & 0&.0 \\
	\multicolumn{6}{c}{
		\begin{tabular} {c c c c c}
			Ref. & $a$ (keV) & $b$ (keV) & $c$ (keV) & $\chi^2$ \\
			\hline
			This Work & 4898.4 (13) & -3651.36 (63) & 235.00 (77) & 28.0\\
			Ref. \cite{2013Lam} & 4894 (1) & -3662 (2) & 243 (2) & 3.0\\
			\hline
		\end{tabular}
	}\\
	\multicolumn{6}{c}{
		\begin{tabular}{l c c}
		Fit & $d$ (keV)& $\chi^{2}$\\
		\hline
		% Quadratic & - & 28\\
		Cubic & $6.7\ (13)$ & -\\
		USDA & $-0.3$ &\\
		USDB & 0.3 &\\
		\end{tabular}
	}\\
	\hline
	\hline
	\multicolumn{6}{l}{$A$ = 21, $J^{\pi} = 1/2^{+}$, $T = 3/2$} \\
	$^{21}$F  & +3/2 & -47&.6 (18)	  & 279&.93 (6)  \\
	$^{21}$Ne  & +1/2 & -5731&.78 (4)  & 9148&.9 (16)  \\ 
	$^{21}$Na  & -1/2 & -2184&.6 (3)  & 9217&.0 (20)	  \\
	$^{21}$Mg  & -3/2 & 10903&.85 (74) \footnotemark[2] & 200&.5 (28) \footnotemark[3]\\
	\multicolumn{6}{c}{
		\begin{tabular} {c c c c c}
			Ref. & $a$ (keV) & $b$ (keV) & $c$ (keV) & $\chi^2$ \\
			\hline
			This Work & 5170.4 (14) & -3633.6 (10) & 220.9 (10) & 9.7\\
			Ref. \cite{2013Lam} & 5171 (10) & -3617 (2) & 217 (2) & 3.5\\
			\hline
		\end{tabular}
	}\\
	\multicolumn{6}{c}{
		\begin{tabular}{l c c}
		Fit & $d$ (keV)& $\chi^{2}$\\
		\hline
		% Quadratic & - & 9.7\\
		Cubic & $-4.4\ (14)$ & -\\
		USDA & $-1.2$ &\\
		USDB & 1.9 &\\
		\end{tabular}
	}\\
	\hline
	\hline
\end{tabularx}
\footnotetext[1]{Average of Refs. \cite{2012Wallace,2010Wrede}}
\footnotetext[2]{Present work}
\footnotetext[3]{Average of Refs. \cite{2004Firestone,2008Diget}}
\end{minipage}
\label{table_IMMEFits}
\end{table}

The TITAN system \cite{2006Dilling} currently consists of three ion traps: an 
RFQ cooler and buncher \cite{2012Brunner}, an 
electron beam ion trap (EBIT) \cite{2012Lapierre} for charge breeding and in-trap decay spectroscopy, and a 
measurement Penning trap (MPET) \cite{2009Brodeur} 
to precisely determine atomic masses. 
We bypassed the EBIT because the required precision could be reached without charge breeding.
In the MPET the mass is determined by measuring an ion's cyclotron frequency \(\nu_{c} = qB/(2 \pi m)\)
via the time-of-flight ion cyclotron resonance technique \cite{1995Konig}.
A typical resonance for $^{21}$Mg is shown in Fig. \ref{fig_20Mgres}.
To eliminate any dependence on the magnetic field the ratio of cyclotron frequencies $r = \nu_{c,ref}/\nu_{c}$ is taken between a well-known reference and the ion of interest. 
The reference ion used was $^{23}$Na.
Because the magnetic field fluctuates in time, due to field decays and temperature or pressure variations, it must be monitored.
This monitoring is achieved by performing a reference measurement 
before and after a measurement of the ion of interest and linearly interpolating to the time when the frequency of the ion of interest was measured.
To further limit these fluctuations, the length of a measurement was limited to approximately one hour.
The atomic mass of the nuclide of interest is then calculated from
\begin{equation}
M=r(M_{ref}-m_{e})+m_{e}
\end{equation}
where $M$ is the atomic mass of a nuclide, $m_{e}$ is the electron
mass and we neglect electron binding energies.
% We follow the procedure outlined in \cite{2010Ettenauer} to account for potential shifts due to ion-ion interactions, and also include the conservative mass dependent shift given in \cite{2009bBrodeur}.
Although no contaminants were observed during the measurements, to be conservative we performed a ``count class'' analysis to account for ion-ion interactions. In the case of $^{20}$Mg, the statistics were too low ($\approx 50$ ions/s) for such an analysis. Instead we follow the procedure in Ref. \cite{2010Ettenauer}, where the difference between the frequency ratio taken with one detected ion and the ratio with all detected ions was added as a systematic.
We also include the conservative mass dependent shift given in Ref. \cite{2009bBrodeur}, which in the cases here amounts to $\approx 10$ ppb.
Table \ref{table_masses} summarizes the measurement results.
Our mass excess of 10903.85 (74)~keV for $^{21}$Mg
agrees well with the tabulated value from AME2012 \cite{2012Wang}. 
The mass excess of 17477.7 (18)~keV for $^{20}$Mg deviates by 81 keV as compared to the
AME2012, which is a shift of 3$\sigma$. In both cases the precision was increased by more than one order of magnitude.

Determining the energy level of an isospin multiplet member relies on knowing both the ground-state and excited-state energies accurately.
For different experimental techniques, the measured excitation energy depends on separation energies, which can change with improved mass measurements.
New mass measurements of the ground states of $^{20}$Na \cite{2010Wrede} and $^{19}$Ne \cite{2008Geithner} led to an improved proton separation energy value for $^{20}$Na of 2190.1(11) keV.
This value is required to derive the excitation energy of the $J^{\pi} = 0^{+},\ T = 2$ state in $^{20}$Na.
Combining a new measurement of the excitation energy \cite{2012Wallace} with the value compiled in \cite{2010Wrede}, an averaged value of 6524.0(98) keV is obtained, a result that is shifted by 10 keV relative to the tabulated value \cite{2013nndc}.
In $^{21}$Mg, a new measurement of the $J^{\pi} = 1/2^{+}$ state was completed \cite{2008Diget}, which, when averaged with the NNDC \cite{2004Firestone} value, yields 200.5(28) keV.
Both of these new values are included in the following analysis.

Table \ref{table_IMMEFits} summarizes the
fit results of the quadratic, and higher order forms of the IMME for the
$A=20$ and $21$ multiplets.
For each multiplet the $\chi^{2}$ of the fit greatly increased, as compared to the tabulated values \cite{2013Lam}.
The most uncertain member of the $A=20$ multiplet is now $^{20}$Na, with nearly all of the uncertainty originating from the excitation energy of the $T = 2$ state.
The best fit is obtained when a cubic term is included, resulting in $d = 2.8 (1.1)$ keV, and $\chi^2 = 3.7$, a result that is an order of magnitude larger than the literature $\chi^2$ value of 0.2 \cite{2013Lam}.
For the $T=3/2$, $A=21$ multiplets, the $\chi^2$ for a quadratic fit have increased to 28 and 9.7 for the $J^{\pi} = 5/2^{+}$ and $1/2^{+}$, respectively, as compared to the literature $\chi^2$ values of 3.0 and 3.5 \cite{2013Lam}.
The IMME clearly fails in both of the $A = 21$ multiplets.
Large cubic terms are required for both multiplets, with $d = 6.7(13)$ for the $J^{\pi} = 5/2^{+}$ multiplet and $d = -4.4(14)$ for the $J^{\pi} = 1/2^{+}$ multiplet.

Exploring the impact of isospin-symmetry breaking on nuclear structure
is an exciting field of research, as it relates to the
symmetries of QCD and their breaking. The $sd$ shell is particularly
interesting because it can be accessed by phenomenological and ab initio
methods. The phenomenological isospin-symmetric USD
Hamiltonians~\cite{usda} generally reproduce data very well throughout
the $sd$ shell, but ultimately need to be supplemented with an isospin
non-conserving (INC) part~\cite{ormand}. In addition, there are
valence-space calculations based on chiral two-nucleon (NN) and
three-nucleon (3N) forces, without phenomenological adjustments. The
resulting $sd$ shell Hamiltonians are inherently isospin asymmetric
and have successfully described proton- and neutron-rich
systems~\cite{Oxygen2,protonrich}, but it is still an open question
how well they work in systems with both proton and neutron valence
degrees of freedom. Therefore the current measurements provide
valuable new tests of these methods.

We first calculated the IMME in the $sd$ shell with the USDA and USDB isospin-conserving 
Hamiltonians \cite{usda}, supplemented with the INC Hamiltonian of Ref.~\cite{ormand}.  The 
results for the $A=20$ and $21$ $d$ and $e$ coefficients are presented in 
Tab.~\ref{table_IMMEFits}.  
For $A = 20$, the USDA value for $e$, which comes from mixing of states with similar energy 
but different isospin in $^{20}$F, $^{20}$Ne, and $^{20}$Na, agrees with
experiment only when $e$ is also included in the IMME fit.  Here, the largest mixing comes 
from a pair of close-lying $J^{\pi} = 0^+$, $T=0,2$ states in $^{20}$Ne.
With the USDB Hamiltonian, these two states are nearly degenerate
resulting in an uncertainty of the energy of the good isospin states that is too large to give a meaningful result. The calculated $d$ 
term, on the other hand, comes from mixing in $^{20}$F and $^{20}$Na.  
% With USDB the $J^{\pi} = 0^+$, $T=0,2$ levels in these nuclei are well separated, leading to a small energy-mixing shift.  
% There are, however, many $T=1$ levels near $T=2$ that are known 
% experimentally but have unidentified spin \cite{2013nndc}.  If one of these states were an 
% intruder $J^{\pi} = 0^+$ state, there could be enough mixing with the $T=2$ state to yield the 
% large observed $d$ value.
With the USDB, the $J^{\pi} = 0^{+}$, $T = 1$ levels in these nuclei are well separated from the $T = 2$ isobaric analog state (IAS), leading to a small energy-mixing shift and hence a too small $d$ value.

For the $A=21$ systems, the USD values of $d$ also do not agree with experiment.
The non-zero values come from mixing of the $T=3/2$ states with close-lying $T=1/2$ states in $^{21}$Ne and $^{21}$Na.
For instance, the largest shift in the $J^{\pi} = 5/2^{+}$ multiplet is due to a $T = 1/2$ state in $^{21}$Ne, which for the USDA lies 372~keV below the IAS, instead of the 50~keV necessary to reproduce $d = 6.7$~keV. Experimentally, several $T = 1/2$ states with unidentified spin lie around the IAS \cite{2013nndc}. This illustrates the challenge in obtaining accurate calculations capable of describing the new experimental findings.
% The largest shift in the $J^{\pi} = 5/2^+$ multiplet is due to a $T=1/2$ state in $^{21}$Ne.
% In order to achieve $d$ = 6.7 keV, this state would have to lie about 50 keV below the $J^{\pi} = 5/2^+$, $T=3/2$ isobaric analogue state (IAS).  Since there are experimental levels in this region with unidentified spin, 
% \cite{2013nndc},  it is possible that the experimental results can be explained by 
% mixing with nearby $T=1/2$ states. In the USDA, however, this state is instead 372 keV 
% below the $T=3/2$ IAS, illustrating the difficulty in obtaining accurate calculations capable of 
% describing the new experimental findings.

In addition we calculated the properties of the $A=20$ and $21$ multiplets from 
valence-space Hamiltonians constructed within the framework of many-body perturbation 
theory \cite{mbpt}, based on low-momentum \cite{Vlowk} NN and 3N forces derived from 
chiral effective field theory \cite{2009Epelbaum}, without empirical adjustments. These 
isospin-asymmetric Hamiltonians describe ground- and excited-state properties in neutron-rich 
oxygen isotopes \cite{Oxygen,Oxygen2,O26} and proton-rich $N=8$ isotones 
\cite{protonrich}. Here we use the valence-space Hamiltonians of Refs.~\cite{Oxygen2} and
Ref.~\cite{protonrich} for the neutron-neutron and proton-proton parts, respectively, and 
include for the first time valence-space proton-neutron interactions. The ground-state 
energies of $^{20,21}$Mg are shown in Table \ref{table_binding}. The calculated 
ground-state energy of $^{20}$Mg is in very good agreement with experiment, while 
$^{21}$Mg, with one neutron above the closed $N = 8$ shell, is overbound by 1.6 MeV. For 
the $A=20$ multiplet, the $d$ coefficient is found to be $-18$ keV, i.e. giving 
isospin-symmetry breaking larger than in experiment.  As $T_z$ increases, other members 
of the $A = 21, T=3/2$ multiplet become less overbound than $^{21}$Mg ($^{21}$F is only 
0.8 MeV overbound). This, however, also results in larger cubic terms for the $A = 21$ 
multiplets ($d = -38$ keV for $A = 21$, $J^{\pi} = 5/2^{+}$).  Therefore, the new experimental 
findings cannot be described with these Hamiltonians, but they nonetheless provide a 
promising first step towards understanding isospin-symmetry breaking based on 
electromagnetic and strong interactions. 

\begin{table}
\caption{Experimental and calculated ground-state energies (in MeV) of 
$^{20,21}$Mg with respect to $^{16}$O. USDA/B results include the INC 
Hamiltonian discussed in the text.}
\begin{tabular}{c r@{}l r@{}l r@{}l r@{}l}
	\hline
	\hline
	Nuclide & \multicolumn{2}{c}{Exp.} & \multicolumn{2}{c}{USDA} & \multicolumn{2}{c}{USDB} & \multicolumn{2}{c}{NN + 3N} \\
	\hline 
	$^{20}$Mg & $-6$&$.94$ & $-6$&$.71$ & $-6$&$.83$ & $-6$&$.89$ \\
	$^{21}$Mg & $-21$&$.59$ & $-21$&$.79$ & $-21$&$.81$ & $-23$&$.18$ \\
	\hline 
	\hline
\end{tabular}
\label{table_binding}
\end{table}

In summary, we have performed the first direct mass measurement of $^{20,21}$Mg using the 
TITAN Penning trap, making $^{20}$Mg the most exotic proton-rich nuclide to be 
measured in a Penning trap. The new mass of $^{20}$Mg is 15 times more precise 
and deviates from the AME12 value by 3$\sigma$, while the new mass of 
$^{21}$Mg agrees with the AME12 but is 22 times more precise. A quadratic fit of 
the $A=20$ IMME results in a $\chi^2$ of 10.4, reducing to 3.7 or 9.9 with the 
inclusion of cubic or quartic terms, respectively. The increased precision of the 
$^{21}$Mg mass now gives, for the $5/2^{+}$ multiplet, $\chi^2$ = 28 with a quadratic IMME, making the 
$A = 20$ and $21$ multiplets new members of the $sd$ shell to present 
strong deviations from the quadratic form of the IMME. In both cases shell-model 
calculations are presently unable to predict the required cubic $d$ coefficients. Further 
Penning trap measurements of members of other multiplets will be
valuable to better test some of the proposed mechanisms for cubic and quartic 
terms. With the on-line use of the IG-LIS, such exotic nuclei will be available
contaminant free, paving the way for new experiments far from stability.

\acknowledgments

This work was supported by the Natural Sciences and Engineering
Research Council of Canada (NSERC). A. T. G. acknowledges support from the NSERC CGS-D program
and A. L. from the DFG under Grant No. FR601/3-1.
This work was also supported by the BMBF under Contract No.~06DA70471, the 
DFG through Grant SFB 634, the Helmholtz Association through the Helmholtz 
Alliance Program, contract HA216/EMMI ``Extremes of Density and Temperature: 
Cosmic Matter in the Laboratory''. Computations were performed with an allocation 
of advanced computing resources at the J\"ulich Supercomputing Center.
We acknowledge support from NSF grant PHY-1068217 and the NSF REU program at MSU.

\end{document}